\documentclass[11pt,reprint,amsmath,amssymb,
 nofootinbib, aps,prl]{revtex4-2}
 
\usepackage{geometry}
\usepackage{amsmath,amsthm,amssymb, mathtools,mathrsfs}
\usepackage{physics}
\usepackage{amsfonts, latexsym, amssymb}
\usepackage{bbm, dsfont}
\usepackage{caption}
\usepackage{graphicx}
\usepackage{float}
\usepackage{url}
\usepackage{xcolor}
\usepackage{hyperref}
\usepackage{tikz}
\usetikzlibrary{decorations.pathmorphing,cd,decorations.markings,calc,fadings}
\hypersetup{
    colorlinks=true,
    linkcolor=blue,
    filecolor=magenta,      
    urlcolor=cyan,
}
\usepackage{enumerate}
\usepackage{bm}% bold math

\newcommand{\be}{\begin{equation}} \newcommand{\ee}{\end{equation}}
\newcommand{\ben}{\begin{equation*}} \newcommand{\een}{\end{equation*}}

\newcommand{\bea}{\begin{equation} \begin{aligned}} \newcommand{\eea}{\end{aligned} \end{equation}}

\def\repa{\raise4pt\hbox{$\square$}\mkern-14mu\raise-4pt\hbox{$\square$}}
\def\repab{\overline{\raise4pt\hbox{$\square$}\mkern-14mu\raise-4pt\hbox{$\square$}\mkern-1mu}}

\begin{document}

\title{
Emergent non-invertible symmetries in $\mathcal{N}=4$ Super-Yang-Mills theory
}
\author{Orr Sela}
\affiliation{Mani L. Bhaumik Institute for Theoretical Physics, Department of Physics and Astronomy, University of California, Los Angeles, CA 90095, USA}

\date{\today}

\begin{abstract}
One of the simplest examples of non-invertible symmetries in higher dimensions appears in 4d Maxwell theory, where its $SL(2,\mathbb{Z})$ duality group can be combined with gauging subgroups of its electric and magnetic 1-form symmetries to yield such defects at many different values of the coupling. Even though $\mathcal{N}=4$ Super-Yang-Mills (SYM) theory also has an $SL(2,\mathbb{Z})$ duality group, it only seems to share two types of such non-invertible defects with Maxwell theory (known as duality and triality defects). Motivated by this apparent difference, we begin our investigation of the fate of these symmetries by studying the case of 4d $\mathcal{N}=4$ $U(1)$ gauge theory which contains Maxwell theory in its content. Surprisingly, we find that the non-invertible defects of Maxwell theory give rise, when combined with the standard $U(1)$ symmetry acting on the free fermions, to defects which act on local operators as elements of the $U(1)$ outer-automorphism of the $\mathcal{N}=4$ superconformal algebra, an operation that was referred to in the past as the "bonus symmetry". Turning to the nonabelian case of $\mathcal{N}=4$ SYM, the bonus symmetry is not an exact symmetry of the theory but is known to emerge at the supergravity limit. Based on this observation we study this limit and show that if it is taken in a certain way, non-invertible defects that realize different elements of the bonus symmetry emerge as approximate symmetries, in analogy to the abelian case.
\end{abstract}

\maketitle

\section{Introduction}

A recent exciting development in quantum field theory is the understanding that symmetries are represented by topological defects with properties that might go beyond the traditional notion of a symmetry. Non-invertible symmetries, in particular, correspond to defects with exotic non-group-like fusion rules and by now have been identified and investigated in diverse setups in different areas of physics (see \cite{Schafer-Nameki:2023jdn,Shao:2023gho} for reviews). 

One of the simplest instances of such non-invertible symmetries in higher dimensions appears in 4d Maxwell theory. As shown in \cite{Choi:2021kmx,Choi:2022zal}, when its $SL(2,\mathbb{Z})$ duality is combined with gauging a $\mathbb{Z}_{N}^{(1)}$ subgroup of its electric 1-form symmetry (with a Dirichlet boundary condition for the $\mathbb{Z}_{N}^{(1)}$ 2-form gauge field, and possibly with stacking an SPT phase for it), one is able to construct non-invertible defects, known as duality and triality defects, at special values of the complexified coupling $\tau$. Later \cite{Niro:2022ctq} (see also \cite{Cordova:2023ent}), employing the observation that gauging a subgroup of the electric or magnetic 1-form symmetry results again in Maxwell theory but with a different coupling, the $SL(2,\mathbb{Z})$ duality was combined with gauging such subgroups (in a non-anomalous way) to yield an $SL(2,\mathbb{Q})$ operation on the theory which acts in an analogous way to $SL(2,\mathbb{Z})$ (this construction will be reviewed in the next section). This, in turn, enabled to find a new non-invertible defect at any value of $\tau$ that is fixed by an element of $SL(2,\mathbb{Q})$ (which acts on $\tau$ in the standard way by a fractional transformation), thereby generalizing the results of \cite{Choi:2021kmx,Choi:2022zal}. 

It is natural at this point to ask what part of this construction has a counterpart in non-abelian theories. Due to the central role played by $SL(2,\mathbb{Z})$ duality, the natural candidate to examine, which will also be the main focus of this note, is $\mathcal{N}=4$ Super-Yang-Mills (SYM) theory. Considering for concreteness the theory with $SU(N)$ gauge group, even though it shares with Maxwell theory an $SL(2,\mathbb{Z})$ duality group, it only has an electric 1-form symmetry which is $\mathbb{Z}_{N}^{(1)}$ and whose gauging changes the global structure of the theory and cannot be associated with an action on the coupling $\tau$. As a result, we do not seem to have an $SL(2,\mathbb{Q})$ operation as in Maxwell theory, and a non-invertible defect of the type we discuss only has the chance of being found at values of $\tau$ that are fixed by $SL(2,\mathbb{Z})$.\footnote{Notice that we only consider here defects involving $SL(2,\mathbb{Z})$ duality operations. Other non-invertible defects might exist at different values of $\tau$, such as condensation defects \cite{Roumpedakis:2022aik,Gaiotto:2019xmp} which are found at any $\tau$.} Indeed, previous works \cite{Choi:2022zal,Kaidi:2022uux} only found the duality and triality defects in $\mathcal{N}=4$ SYM, with the additional defects discussed in \cite{Niro:2022ctq} being absent. 

Here, in order to investigate this apparent difference between Maxwell and $\mathcal{N}=4$ SYM theories in a more systematic way, and to try to extract a clue that will guide us towards the fate of these additional symmetries in $\mathcal{N}=4$ SYM, we begin by examining more closely a theory which is in a sense the intersection of these two: 4d $\mathcal{N}=4$ abelian $U(1)$ gauge theory. This theory both has an $\mathcal{N}=4$ superconformal algebra and contains free Maxwell theory in its content. Surprisingly, we will find that the additional symmetries of \cite{Niro:2022ctq}, when combined with the standard $U(1)$ symmetry acting on the free fermions in the theory, realize at different values of the coupling different elements of a non-invertible $U(1)$ $R$-symmetry that acts on the local operators as the $U(1)$ outer-automorphism
of the $\mathcal{N}=4$ superconformal algebra. Moreover, since correlators of local operators do not depend on the coupling in a nontrivial way in this theory (i.e. one can normalize the operators such that the correlators are coupling independent), they will satisfy the selection rules 
of the entire $U(1)$ at any value of the coupling.

Apriori, the $U(1)$ outer-automorphism of the $\mathcal{N}=4$ superconformal algebra may or may not be a symmetry of the theory. This question was discussed in detail in the past in \cite{Intriligator:1998ig,Intriligator:1999ff} (see also \cite{SCHWARZ1983301,HOWE1984181,SCHWARZ1983269,Green:1997tv}), where this potential symmetry was referred to as the {\it bonus symmetry}. It was demonstrated that in the abelian theory we consider, the various local operators have definite charges under this $U(1)$ and that it is respected by the equations of motion and supersymmetry transformations. However, the nature of this symmetry seemed to be mysterious as it was clearly not a symmetry of the Lagrangian and the field strength appeared to be charged under it. One of the observations made in this note is therefore the identification in modern terms of the bonus symmetry of the $U(1)$ gauge theory as a non-invertible symmetry, with different elements realized at different values of the coupling. These elements, in turn, mainly correspond to the defects discovered in \cite{Niro:2022ctq}, and while their action on local operators is the one identified in the past in \cite{Intriligator:1998ig,Intriligator:1999ff}, their action on line operators is highly nontrivial \cite{Niro:2022ctq}. 
 
Once we have identified the non-invertible defects of Maxwell theory discussed above as the key ingredient giving rise to the bonus symmetry of the $\mathcal{N}=4$ theory with $U(1)$ gauge group, it is time to turn to the nonabelian theory and use the bonus symmetry of its algebra as our guide in searching for new non-invertible symmetries analogous to those of Maxwell theory. Unlike the abelian case, in $\mathcal{N}=4$ SYM the bonus symmetry is not an exact symmetry of the theory \cite{Intriligator:1998ig,Intriligator:1999ff}. This is indeed consistent with the fact that only the duality and triality defects have been identified as exactly-topological defects in the past. However, as discussed in detail in \cite{Intriligator:1998ig}, based on holographic duality the bonus symmetry is expected to emerge as an approximate symmetry in the limit where the gravity dual of $\mathcal{N}=4$ SYM is approximated by type IIB supergravity (i.e. when both $N$ and $g_{YM}^2N$ are very large). This emergence, in turn, follows from the enhancement of the $SL(2,\mathbb{Z})$ duality symmetry of type IIB string theory to $SL(2,\mathbb{R})$ in this supergravity limit (at least when the fields are not
treated as quantized), of which the $U(1)$ bonus symmetry is the maximal compact subgroup. This observation then suggests that the place in $\mathcal{N}=4$ SYM in which we should look for new defects analogous to those of Maxwell theory is exactly at this limit. 

In order to do it, we will first investigate the global structure of $\mathcal{N}=4$ SYM at the large-$N$ limit (without specifying the value of the 't Hooft coupling $\lambda$). Focusing on a gauge group of the form $SU(N^2)$ we will find, following an approach closely related to the one recently discussed in \cite{Putrov:2022pua}, that the large-$N$ limit can be taken in such a way that the $\mathbb{Z}_{N^2}^{(1)}$ 1-form symmetry turns into $\mathbb{\widehat{Q}}^{(1)}$ at the limit, where $\mathbb{\widehat{Q}}$ is called the ring of profinite rational numbers (which will be defined below) and contains the rationals $\mathbb{Q}$ as a subring. As we will show, this results in the $SL(2,\mathbb{Z}_{N^2})$ group, associated with the operations of gauging the $\mathbb{Z}_{N^2}^{(1)}$ 1-form symmetry and stacking an SPT phase at finite $N$, turning into $SL(2,\mathbb{\widehat{Q}})$ at this limit. This will then allow us to identity at the supergravity regime (i.e. when $\lambda$ is taken to be very large) an $SL(2,\mathbb{Q})\subset SL(2,\mathbb{R})$ duality group, consistent with this structure, which when combined with the $SL(2,\mathbb{Q})\subset SL(2,\mathbb{\widehat{Q}})$ operations yields new non-invertible defects. We obtain a new such defect at any value of $\lambda$ (or $\tau$) that is both in the supergravity regime and is fixed by an element of $SL(2,\mathbb{Q})$, analogously to the situation in Maxwell theory.\footnote{Note that unlike $SL(2,\mathbb{Z})$, $SL(2,\mathbb{Q})$ can fix a $\tau$ that corresponds to an arbitrarily small gauge coupling (or $1/\textrm{Im}(\tau)$).} In addition, defects at different such
couplings realize different elements of the bonus symmetry, as in the case of the $\mathcal{N}=4$ $U(1)$ gauge theory.

\section{$\mathcal{N}=4$ abelian $U(1)$ gauge theory}

In this section we would like to investigate the way in which the non-invertible defects of Maxwell theory that include $SL(2,\mathbb{Z})$ duality transformations appear in the 4d $\mathcal{N}=4$ theory with $U(1)$ gauge group. As discussed in the previous section, we are mainly interested in the relation between these defects and the bonus (or outerautomorphism) symmetry of the $\mathcal{N}=4$ theory. 

Before beginning with reviewing the non-invertible defects of Maxwell theory, let us take a small detour and discuss some hints for them in the classical theory. Defining the self dual and anti self dual field strengths $F_{mn}^{\pm}=\frac{1}{2}(F_{mn}\pm\frac{1}{2}\epsilon_{mnpq}F^{pq})$, the Lagrangian in Euclidean signature takes the form 
\begin{equation}
    \mathcal{L}=\frac{i}{8\pi}\left(\overline{\tau}F_{mn}^{+}F^{+mn}-\tau F_{mn}^{-}F^{-mn}\right).
\end{equation}
Clearly, there does not seem to be any (0-form) symmetry beyond charge conjugation that will act on $F_{mn}^{\pm}$ (and not on $\tau$) while leaving the Lagrangian invariant. However, since our modern understanding of symmetries in quantum field theory is in terms of topological defects, we would actually like to look for symmetries of the stress tensor instead of the Lagrangian. This is because a topological defect necessarily acts trivially on the stress tensor (alternatively, the displacement operator on the defect vanishes), but does not necessarily correspond to a symmetry of the Lagrangian. Let us therefore look for classical operations on the fields that will leave the stress tensor invariant.\footnote{Note that this is not a sufficient condition for a symmetry in the quantum theory. For example, the $U(1)$ axial symmetry of QCD leaves the stress tensor invariant but does not correspond to a topological defect in the quantum theory (only a discrete subgroup of it does).} In Maxwell theory, we have 
\begin{equation}
\label{T_Maxwell}
    T_{mn}=\frac{\textrm{Im}\left(\tau\right)}{4\pi}\left(F_{mk}^{+}F_{\quad n}^{-k}+F_{mk}^{-}F_{\quad n}^{+k}\right)
\end{equation}
and we can readily observe that $F^{\pm}\rightarrow e^{\pm i\varphi}F^{\pm}$ is a symmetry of $T_{mn}$. This operation is also a symmetry of the equations of motion, but a bit strange at first sight since the field strength is charged under it. 

To understand if and how this $U(1)$ symmetry can appear in the quantum theory, we notice that its action on $F^{\pm}$ implies that it rotates between $F$ and $i\star F$, or alternatively between the electric and magnetic fields. This suggests that if this symmetry is realized at the quantum level, it involves the $SL(2,\mathbb{Z})$ duality of Maxwell theory. This was indeed shown to be the case in \cite{Niro:2022ctq}, as we now briefly review. 

We begin with the observation that gauging e.g. a $\mathbb{Z}_{p}^{(1)}$ subgroup of the $U(1)^{(1)}_e$ electric 1-form symmetry results in a theory which has the same coupling $\tau$ but with the gauge field $A$ having fluxes $\mathbb{Z}/p$ (and with the charges of Wilson lines being $p\mathbb{Z}$). Writing the gauge field as $A=A'/p$ with $A'$ having fluxes $\mathbb{Z}$, we can then rewrite the theory in terms of $A'$ and end up with the same Lagrangian and gauge-field fluxes we had at the beginning but with coupling $\tau/p^2$ instead of $\tau$. During this total transformation consisting of 1-form gauging plus rescaling, a Wilson line with charge $p$ in the original theory maps to a Wilson line with charge $1$ at the end, while an 't Hooft line with charge $1$ maps to one with charge $p$. We see that we can represent this transformation by the matrix $\textrm{diag}(1/p,p)$ acting in the usual way on the doublet of electric and magnetic 1-form charges $(e,m)$ and by a fractional transformation on the coupling $\tau$, just like the way $SL(2,\mathbb{Z})$ acts. Overall, combining such (non-anomalous) electric and magnetic 1-form gauging with the usual $SL(2,\mathbb{Z})$ duality yields the following $SL(2,\mathbb{Q})$ operation, 
\begin{equation}
    \left(\begin{array}{c}
e\\
m
\end{array}\right)\rightarrow\left(\begin{array}{cc}
q_{1} & q_{2}\\
q_{3} & q_{4}
\end{array}\right)\left(\begin{array}{c}
e\\
m
\end{array}\right)\quad,\quad\tau\rightarrow\frac{q_{1}\tau+q_{2}}{q_{3}\tau+q_{4}}
\end{equation}
with the corresponding matrix being an element of $SL(2,\mathbb{Q})$. Such a matrix can be obtained by the following sequence of operations (with $q_{4}=(1+q_{2}q_{3})/q_{1}$)
\begin{equation}
\label{Matrix_decom}
    \left(\begin{array}{cc}
q_{1} & q_{2}\\
q_{3} & q_{4}
\end{array}\right)=\left(\begin{array}{cc}
1 & 0\\
\frac{q_{3}}{q_{1}} & 1
\end{array}\right)\left(\begin{array}{cc}
1 & q_{1}q_{2}\\
0 & 1
\end{array}\right)\left(\begin{array}{cc}
q_{1} & 0\\
0 & q_{1}^{-1}
\end{array}\right),
\end{equation}
where the rightmost matrix corresponds to gauging subgroups of the electric and magnetic 1-form symmetries, the middle matrix to shifting the $\theta$-term by a rational number which can be obtained by a $CTST$-type magnetic gauging\footnote{Alternatively, this matrix can be obtained by combining $T$-duality with gauging a subgroup of the electric and magnetic 1-form symmetries \cite{Niro:2022ctq}.} (where $C$ denotes charge conjugation, $S$ gauging a subgroup of the magnetic 1-form symmetry and $T$ stacking an SPT phase) \cite{Choi:2022jqy}, and the left matrix can be obtained from the middle one by combining it with an $S$-duality operation and its inverse, see \cite{Niro:2022ctq} for more details. It is important to comment that the operations in \eqref{Matrix_decom} are performed sequentially such that in each step we return to Maxwell theory with gauge fluxes $\mathbb{Z}$ and some coupling, such that there is no tension between these operations and the anomaly between the electric and magnetic 1-form symmetries. 

At this point one can easily check that every coupling of the form 
\begin{equation}
\label{Fixed_tau}
    \tau=\frac{q_{1}-q_{4}+i\sqrt{4-\left(q_{1}+q_{4}\right)^{2}}}{2q_{3}}
\end{equation}
with $q_1$, $q_2$ and $q_4$ any rationals satisfying $2>q_{1}+q_{4}$ and $q_{3}\neq0$ (such that the coupling $g$ is kept real) is invariant under the corresponding $SL(2,\mathbb{Q})$ element, with $q_{2}=(q_{1}q_{4}-1)/q_{3}$. We therefore see that topological defects corresponding to different elements of $SL(2,\mathbb{Q})$ are realized at different values of the coupling.\footnote{Let us comment that each such element of $SL(2,\mathbb{Q})$ can be realized by more than one defect, differing by condensation defects.} Moreover, such defects rotate between the electric and magnetic fields with different rotation angles, and since correlators of local operators do not depend on $\tau$ in a nontrivial way in this theory, such correlators will respect the selection rules of the entire $U(1)$ operation discussed below Eq. \eqref{T_Maxwell}. Notice, however, that this $U(1)$ is not really a symmetry of the theory and is not associated with a topological operator. For more details on this construction, as well as explicit Lagrangian descriptions of such $SL(2,\mathbb{Q})$ defects and a demonstration of their non-invertibility, see \cite{Niro:2022ctq}. 

Turning now to the $\mathcal{N}=4$ $U(1)$ gauge theory, the status of its $U(1)$ bonus symmetry \cite{Intriligator:1998ig} is very similar to that of the $U(1)$ "symmetry" of Maxwell theory we discussed above. To see it, let us begin with the basic description of the theory. We consider a free theory consisting of a $U(1)$ gauge field with field strength $F_{(\alpha\beta)}$, fermions $\psi_{I\alpha}$, $\overline{\psi}_{\dot{\alpha}}^{I}$ and real scalars $\phi_{[IJ]}$, where $I$ is the index of the fundamental representation of $SU(4)_R$ and $\alpha$, $\dot{\alpha}$ the usual indices of $SU(2)_{L,R}$. The Lagrangian is simply given by the sum of the kinetic terms of the fields, with the exactly marginal coupling $\tau$ (and $\overline{\tau}$) an overall factor. Since this theory contains Maxwell theory, it also contains the defects we discussed above, realizing when acting on local operators different elements of a $U(1)$ symmetry which we will denote by $U(1)_F$. As its name suggests, among the basic fields of the theory only the field strength is charged under $U(1)_F$ (with charge $1$ for $F_{(\alpha\beta)}$ and $-1$ for $\overline{F}_{(\dot{\alpha}\dot{\beta})}$). In addition, there is a $U(1)_\psi$ symmetry under which the free fermions $\psi_{I\alpha}$ are charged with charge 1, and a particularly natural combination of these two symmetries is $U(1)_Y=-U(1)_{\psi}-2U(1)_{F}$ under which the supercurrent has a well-defined charge of $-1$ (indeed, this will be the charge of terms like $F_{(\alpha\beta)}\overline{\psi}_{\dot{\alpha}}^{I}$ and $\partial_{\alpha\dot{\alpha}}\phi^{[IJ]}\psi_{J\beta}$ in it). Since the supercharges are charged under $U(1)_Y$, it is an $R$-symmetry, and in fact this is exactly the bonus symmetry as defined in \cite{Intriligator:1998ig}. We therefore see that in this $\mathcal{N}=4$ $U(1)$ gauge theory, the bonus symmetry appears as an ordinary 0-form $U(1)$ symmetry at the level of local correlators, but is in fact given by (in general) non-invertible defects realizing at different values of the coupling different elements of it. 

Motivated by this link between the non-invertible defects in Maxwell theory associated with its $SL(2,\mathbb{Q})$ operation and the bonus symmetry of the $\mathcal{N}=4$ $U(1)$ gauge theory, we continue in the next section to non-abelian $\mathcal{N}=4$ theories with the aim of using their bonus symmetry as a guide for finding new analogous defects.

\section{$\mathcal{N}=4$ Super-Yang-Mills Theory}

As discussed in the introduction, the bonus symmetry is not an exact symmetry of $\mathcal{N}=4$ SYM, in the sense that the corresponding selection rules are clearly violated in general by local correlators. However, in the supergravity limit of the theory (when both $N$ and $g_{YM}^2N$ are very large) we obtain a description in terms of type IIB supergravity on $AdS_{5}\times S^{5}$, and when the fields are not treated as quantized there is a well-known $SL(2,\mathbb{R})$ symmetry acting on them (enhancing the standard $SL(2,\mathbb{Z})$ duality of IIB string theory). The bonus symmetry then emerges as the $U(1)$ subgroup of this $SL(2,\mathbb{R})$ that fixes a given value of the coupling $\tau$, and the corresponding selection rules are expected to be satisfied by correlators of local operators that can be computed using the supergravity approximation \cite{Intriligator:1998ig}. 

This suggests that defects analogous to the ones discussed in the previous section might emerge as approximate symmetries in the supergravity limit. For this to be the case, the possible global structures of the theory at this limit should allow for transformations analogous to the $SL(2,\mathbb{Q})$ of Maxwell theory. As we will show, such a description can indeed be obtained if the large-$N$ limit is taken in a certain way. 

Let us begin with recalling the case of finite $N$, considering for concreteness the gauge group $SU(N)$. Here there is an electric $\mathbb{Z}_{N}^{(1)}$ 1-form symmetry associated with the $\mathbb{Z}_{N}$ center of $SU(N)$, acting on Wilson lines according to the $N$-ality of their representation. Gauging this 1-form symmetry or a subgroup of it, possibly with stacking an SPT phase, changes the spectrum of line operators in the theory and correspondingly also the 1-form symmetry. The way these different global forms are encoded holographically is through the topological theory \cite{Witten:1998wy} (also known as the SymTFT \cite{Gaiotto:2020iye,Apruzzi:2021nmk,Freed:2022qnc} of the theory) 
\begin{equation}
\label{SymTFT_N}
    S=\frac{2\pi}{N}\int_{AdS_{5}}\mathsf{B}_{N}\cup\delta\mathsf{C}_{N}
\end{equation}
obtained near the boundary of $AdS_{5}$, where $\cup$ is the cup product and $\mathsf{B}_{N}$ and $\mathsf{C}_{N}$ are both $\mathbb{Z}_{N}$ 2-cochains. The different global forms then correspond to different topological boundary conditions for the theory \eqref{SymTFT_N}, which in turn are classified by the Lagrangian subgroups of its surface operators\footnote{We here ignore the issue of topological refinement \cite{Chen:2021xuc}.} 
\begin{equation}
\label{Surfaces_N}
    S_{\left(e,m\right)}\left(\sigma\right)=e^{\frac{2\pi ie}{N}\int_{\sigma}\mathsf{B}_{N}}e^{\frac{2\pi im}{N}\int_{\sigma}\mathsf{C}_{N}}\,.\end{equation}
Surfaces that can end on the boundary (that is, the ones in the chosen Lagrangian subgroup) then correspond to the line operators of the boundary theory which are charged under the 1-form symmetry, while the rest of the surfaces give rise to the symmetry generators when pushed to the boundary. 

Let us now examine the large-$N$ limit of this story. Instead of considering a specific value of $N$ that is finite but large (in which case the discussion remains the same as above), we will take the limit in a way that formalizes more precisely the intuition that combinations such as $(e\int_{\sigma}\mathsf{B}_{N}\;\textrm{mod}\:N)/N$ (which appears in the expression for the surface $S_{(e,m)}$, see \eqref{Surfaces_N}) are approximately valued in all of $\mathbb{Q}/\mathbb{Z}$ as $N$ is taken to be very large. In order to do it, let us focus on the $\mathbb{Z}_{N}$ center of $SU(N)$ and begin by discussing two different natural ways of taking such a large-$N$ limit of it (limits of these types have been recently discussed in a physical context in \cite{Putrov:2022pua}). 

The first, usually referred to as the {\it direct limit}, is defined as follows. Consider the family of groups\footnote{We consider in this limit the groups $\frac{1}{N}\mathbb{Z}_{N}$ instead of $\mathbb{Z}_{N}$ in order to make the construction more transparent. Working with $\mathbb{Z}_{N}$ will yield an isomorphic result.} $\{\frac{1}{N}\mathbb{Z}_{N}\}_{N\in\mathbb{N}}$ together with the inclusion homomorphisms $f_{MN}:\frac{1}{M}\mathbb{Z}_{M}\rightarrow\frac{1}{N}\mathbb{Z}_{N}$ between them when $M|N$. The direct limit, denoted by $\underset{\longrightarrow}{\lim}$, is then given in this case by the disjoint union of all the groups but with an equivalence relation that identifies any element $q\in\frac{1}{M}\mathbb{Z}_{M}$ for some $M$ with all its images under the maps $f_{MN}$. Explicitly, we have 
\begin{equation}
    \underset{\longrightarrow}{\lim}\,\frac{1}{N}\mathbb{Z}_{N}=\underset{N\in\mathbb{N}}{\bigsqcup}\,\frac{1}{N}\mathbb{Z}_{N}\not\;\;\;\sim
\end{equation}
where $\sim$ identifies equal rational numbers that appear in different groups. It is easy to see that the result of this limit is simply $\mathbb{Q}/\mathbb{Z}$, corresponding to the fact that it can be written as the union of all cyclic groups. Let us comment that the direct limit is more general than the specific case considered here, and is applicable to categories that include objects which are more general than the groups we examined and with morphisms that are different from the above homomorphisms. We here followed the general construction and applied it to the large-$N$ limit under consideration. 

The second way in which we can take the large-$N$ limit of $\mathbb{Z}_{N}$ is called the {\it inverse limit}, and is the Pontryagin dual of the direct limit we have previously discussed. We consider the family of groups $\{\mathbb{Z}_{N}\}_{N\in\mathbb{N}}$ with homomorphisms $g_{MN}:\mathbb{Z}_{N}\rightarrow\mathbb{Z}_{M}$ for $M|N$ given by the $\textrm{mod}\:M$ map  (note that now the map is from $\mathbb{Z}_{N}$ to $\mathbb{Z}_{M}$ and not vice versa). The inverse limit, denoted by $\underset{\longleftarrow}{\lim}$, is then given by  
\begin{equation}
\label{Zh}
    \underset{\longleftarrow}{\lim}\,\mathbb{Z}_{N}=\left\{ \vec{a}\in\underset{N\in\mathbb{N}}{\prod}\,\mathbb{Z}_{N}\mid a_{M}=a_{N}\;\textrm{mod}\:M\;\;\forall\;\;M|N\right\} \end{equation}
and simply means that any element in the resulting group is specified by the set of its residues modulo $N$ for all $N$, in a way that is consistent with the $\textrm{mod}$ map relating different residues. This group is called the "profinite integers" and is denoted by $\mathbb{\widehat{Z}}$. It is a certain completion of the integers $\mathbb Z$ and contains them as a subgroup. In particular, there are profinite integers which are not ordinary integers. As for the direct limit, the inverse limit can be defined for more general categories and here we have only considered it in the context of the large-$N$ limit we are taking. 

The natural question at this point is what type of limit should be used for the large-$N$ limit of the $\mathbb{Z}_{N}$ center of the $SU(N)$ gauge group. Using the direct limit would mean that the center turns into $\mathbb{Q}/\mathbb{Z}$ at large $N$, implying that the field $\mathsf{B}_N$ in Eq. \eqref{SymTFT_N} (which can be regarded as a background field for the electric 1-form symmetry) would turn into a $\mathbb{Q}/\mathbb{Z}$ 2-cochain $\mathsf{B}_{\mathbb{Q}/\mathbb{Z}}$. The field $\mathsf{C}_{N}$, on the other hand, which corresponds to the dual symmetry would turn into a $\mathbb{\widehat{Z}}$ 2-cochain $\mathsf{C}_{\mathbb{\widehat{Z}}}$ (since $\mathbb{Q}/\mathbb{Z}$ and $\mathbb{\widehat{Z}}$ are Pontryagin duals). This, however, would mean that the $SL(2,\mathbb{Z}_{N})$ symmetry of the action in \eqref{SymTFT_N} that we have at finite $N$ will not be present at the large-$N$ limit. In addition, there is no natural reason for not treating the fields $\mathsf{B}_{N}$ and $\mathsf{C}_{N}$ on equal footing, taking a direct limit for one and the inverse for the other (alternatively, there is no natural way of choosing which of the global forms, $SU(N)$ or $PSU(N)$, have a $\mathbb{Q}/\mathbb{Z}$ symmetry and which a $\mathbb{\widehat{Z}}$ one). In fact, based on tracing the source of different contributions to the path integral at the large-$N$ limit from those of finite but large $N$, one can show that in a sense both types of limits should be used for both fields. Leaving an analysis of the general case for future work, we will here focus on gauge groups of the form $SU(N^2)$ and take the large-$N$ limit in a way that keeps $\mathsf{B}_{N^2}$ and $\mathsf{C}_{N^2}$ on equal footing and such that the $SL(2,\mathbb{Z}_{N^2})$ symmetry is not broken. 

In order to take such a limit of the $\mathbb{Z}_{N^2}$ center, we view it as the following extension\footnote{I would like to thank P. Putrov for suggesting this way of taking the limit to me.} 
\begin{equation}
\label{sequence_N}
0\rightarrow\mathbb{Z}_{N}\rightarrow\mathbb{Z}_{N^{2}}\rightarrow\mathbb{Z}_{N}\rightarrow0
\end{equation}
where the second arrow denotes multiplication by $N$. Defining the bilinear pairing $(a,b)_{N^{2}}=ab/N^{2}\;\textrm{mod}\:1$ in $\mathbb{Z}_{N^2}$ (which is a map $(,)_{N^{2}}:\mathbb{Z}_{N^{2}}\times\mathbb{Z}_{N^{2}}\rightarrow\mathbb{Q}/\mathbb{Z}$), we observe that the first $\mathbb{Z}_{N}$ group in the sequence \eqref{sequence_N} is a Lagrangian subgroup with respect to it, and that the two $\mathbb{Z}_{N}$ groups can be regarded as Pontryagin dual to each other using it since an element $a\in\mathbb{Z}_{N}$ can be understood as $a\in\textrm{Hom}(\mathbb{Z}_{N^{2}}/\mathbb{Z}_{N},\mathbb{Q}/\mathbb{Z})$ with $a(b)=(a,b)_{N^{2}}$ for $b\in\mathbb{Z}_{N^{2}}/\mathbb{Z}_{N}$ (which is well-defined since $\mathbb{Z}_{N}$ is Lagrangian). We can then take the large-$N$ limit of the sequence \eqref{sequence_N} in such a way that the inverse limit is taken for the first $\mathbb{Z}_{N}$ group while the direct one is taken for the other $\mathbb{Z}_{N}$ (i.e. for $\mathbb{Z}_{N^{2}}/\mathbb{Z}_{N}$). This, in turn, is possible due to the fact that the two $\mathbb{Z}_{N}$ groups are Pontryagin dual to each other and by using the property that Pontryagin duality exchanges direct and inverse limits, $\textrm{Hom}(\underset{\longrightarrow}{\lim}\,A_{n},B)=\underset{\longleftarrow}{\lim}\,\textrm{Hom}(A_{n},B)$. We then end up at this large-$N$ limit with the sequence 
\begin{equation}
0\rightarrow\mathbb{\widehat{Z}}\rightarrow\mathbb{\widehat{Q}}\rightarrow\mathbb{Q}/\mathbb{Z}\rightarrow0\,,
\end{equation}
where the extension $\mathbb{Z}_{N^{2}}$ turns at the limit to the extension $\mathbb{\widehat{Q}}$ of $\mathbb{Q}/\mathbb{Z}$ by $\mathbb{\widehat{Z}}$. The group $\mathbb{\widehat{Q}}$ is called the group of profinite rationals, and is defined in an analogous way to the profinite integers in \eqref{Zh}, 
\begin{equation}
    \mathbb{\widehat{Q}}=\left\{ \vec{q}\in\underset{N\in\mathbb{N}}{\prod}\,\left(\mathbb{Q}/N\mathbb{Z}\right)\mid q_{M}=q_{N}\;\textrm{mod}\:M\;\;\forall\;\;M|N\right\}. 
\end{equation}
It is self-dual under Pontryagin duality and has a natural ring structure extending that of $\mathbb{\widehat{Z}}$. Moreover, it includes $\mathbb{Q}$ and $\mathbb{\widehat{Z}}$ as subrings and can be written as $(\mathbb{\widehat{Z}}\oplus\mathbb{Q})/\mathbb{Z}$. 

We therefore obtain that both $\mathsf{B}_{N^2}$ and $\mathsf{C}_{N^2}$ turn at this limit into $\mathbb{\widehat{Q}}$ 2-cochains, and that the symmetry between them is preserved. In order to identify this symmetry in full and find what $SL(2,\mathbb{Z}_{N^2})$ turns into at this limit, we should first find the limit of the action \eqref{SymTFT_N}. We can do it by rewriting this finite-$N$ action using the pairing $(,)_{N^{2}}$ of $\mathbb{Z}_{N^{2}}$ we defined above as 
\begin{equation}
S=2\pi\int_{AdS_{5}}\left(\mathsf{B}_{N^{2}},\delta\mathsf{C}_{N^{2}}\right)_{N^{2}}
\end{equation}
and observing that this pairing turns at the limit we are taking into the pairing $(q_{1},q_{2})_{\infty}=q_{1}q_{2}\;\textrm{mod}\:\mathbb{\widehat{Z}}\in\mathbb{Q}/\mathbb{Z}$ in $\mathbb{\widehat{Q}}$, where $q_{1},q_{2}\in\mathbb{\widehat{Q}}$ and $q_{1}q_{2}$ is their product using the standard ring structure of $\mathbb{\widehat{Q}}$. We therefore find the large-$N$ action 
\begin{equation}
S_{\infty}=2\pi\int_{AdS_{5}}(\mathsf{B}_{\mathbb{\widehat{Q}}},\delta\mathsf{C}_{\mathbb{\widehat{Q}}})_{\infty}
\end{equation}
and can identify an $SL(2,\mathbb{\widehat{Q}})$ symmetry acting on $\mathsf{B}_{\mathbb{\widehat{Q}}}$ and $\mathsf{C}_{\mathbb{\widehat{Q}}}$ as a doublet using the standard ring structure of $\mathbb{\widehat{Q}}$. Let us also comment that the surfaces of the theory, which for finite $N$ are specified by the pair of charges $(e,m)\in\mathbb{Z}_{N^{2}}\times\mathbb{Z}_{N^{2}}$ (see \eqref{Surfaces_N} for the case of $SU(N)$ gauge group), are now specified by a pair $(\mathfrak{e},\mathfrak{m})\in\mathbb{\widehat{Q}}\times\mathbb{\widehat{Q}}$ and take the form 
\begin{equation}
S_{\left(\mathfrak{e},\mathfrak{m}\right)}\left(\sigma\right)=e^{2\pi i\int_{\sigma}(\mathfrak{e},\mathsf{B}_{\mathbb{\widehat{Q}}})_{\infty}}e^{2\pi i\int_{\sigma}(\mathfrak{m},\mathsf{C}_{\mathbb{\widehat{Q}}})_{\infty}}\,.
\end{equation}

We have found that the group $SL(2,\mathbb{Z}_{N^2})$, corresponding to gauging the 1-form symmetry and stacking an SPT phase at finite $N$, turns into $SL(2,\mathbb{\widehat{Q}})$ at the large-$N$ limit we are considering. Notice, however, that the duality group of $\mathcal{N}=4$ SYM, which acts also on the rest of the theory (e.g. on the coupling $\tau$), is still $SL(2,\mathbb{Z})$ for general values of the 't Hooft coupling $\lambda$. At this point we are making use of the supergravity approximation of the string-theory dual of $\mathcal{N}=4$ SYM which is valid when $\lambda$ is taken to be large, and of its $SL(2,\mathbb{R})$ enhanced symmetry when the fields are treated as classical, to identify an $\ensuremath{SL(2,\mathbb{Q})}\subset\ensuremath{SL(2,\mathbb{R})}$ duality group which is consistent at the quantum level with the global structures (or charge quantization) we have found at the large-$N$ limit we described. Therefore, taking the large-$N$ limit as detailed above and the 't Hooft coupling to be large, we expect to have an $\ensuremath{SL(2,\mathbb{Q})}$ duality group\footnote{Note that we only consider those elements of $\ensuremath{SL(2,\mathbb{Q})}$ which keep $\lambda$ in the supergravity regime.} with elements that leave values of $\lambda$ corresponding to a $\tau$ of the form \eqref{Fixed_tau} invariant. Performing such a duality operation in half-space and accompanying it with the corresponding element of $\ensuremath{SL(2,\mathbb{Q})}\subset\ensuremath{SL(2,\mathbb{\widehat{Q}})}$ that brings the global structure to its original form, we obtain a non-invertible topological defect in the original theory.\footnote{This is similar to the construction of the duality and triality defects. For example, the duality defect at $\tau=i$ is obtained by performing $S$-duality in half-space which is then accompanied by gauging the 1-form symmetry (or performing an $S$ operation in the modular group of gauging and stacking SPT phases) in the same half-space.} Different such defects, which are realized at different values of $\lambda$, implement different elements (corresponding to certain rotation angles) of the $U(1)$ bonus symmetry of $\mathcal{N}=4$ SYM, in analogy to the case of the $\mathcal{N}=4$ $U(1)$ gauge theory discussed in the previous section. Of course, since such defects correspond to approximate symmetries at the supergravity regime which become closer to be topological as $\lambda$ is taken to be larger, we can also consider at each $\lambda$ $SL(2,\mathbb{Q})$ duality elements that only approximately fix it (to the same accuracy).\footnote{Such elements correspond to topological interfaces between two theories with couplings that are almost but not exactly equal. In order to get a defect in the original theory, such an interface has to be accompanied with another interface which changes the coupling back to its original value. These interfaces are only slightly-nontopological and have a very small displacement operator, and therefore result (when combined with the other topological interface) in defects that are similarly almost topological and correspond to approximate symmetries.}

\paragraph{Acknowledgements} I would like to thank O. Bergman, T. Dumitrescu, M. Gutperle, P.-S. Hsin, K. Intriligator, T. Jacobson, P. Kraus, P. Niro, K. Roumpedakis, G. Zafrir and Y. Zheng for useful discussions, and especially to P. Putrov for illuminating discussions about the large-$N$ limit discussed in this paper. I would also like to thank P.-S. Hsin, P. Niro, P. Putrov and K. Roumpedakis for comments on the manuscript. This work was supported by the Mani L. Bhaumik Institute for Theoretical Physics at UCLA.

\bibliographystyle{JHEP}
\bibliography{refs}

%\pagebreak
%\clearpage
%\widetext\begin{center}
%\textbf{\large Supplemental Materials}
%\end{center}

%////////////

%\section{////////////}
%////////////

\end{document}